\documentclass[epj]{svjour}

\usepackage{graphics,graphicx}
\usepackage{amssymb,amsmath}
\usepackage{times,latexsym}

\begin{document}
\title{Model of mobile agents for sexual interactions networks}
\author{Marta C.~Gonz\'alez\inst{1} \and 
        Pedro G.~Lind\inst{1,2} \and 
        Hans J.~Herrmann\inst{1,3}}
\institute{Institute for Computational Physics, 
           Universit\"at Stuttgart, Pfaffenwaldring 27, 
           D-70569 Stuttgart, Germany \and
           Centro de F\'{\i}sica Te\'orica e Computacional, 
           Av.~Prof.~Gama Pinto 2,
           1649-003 Lisbon, Portugal \and
           Departamento de F\'{\i}sica, Universidade Federal do
           Cear\'a, 60451-970 Fortaleza, Brazil}
\date{Received: date / Revised version: date}
\abstract{
We present a novel model to simulate real social networks of complex
interactions, based in a granular system of colliding particles
(agents). 
The network is build by keeping track of the collisions and evolves in 
time with correlations which emerge due to the mobility of the agents.
Therefore, statistical features are a consequence only of local
collisions among its individual agents. 
Agent dynamics is realized by an event-driven algorithm of collisions
where energy is gained as opposed to granular systems which have
dissipation. 
The model reproduces empirical data from networks of sexual
interactions, not previously obtained with other approaches. 
\PACS{{89.65.-s}{Social and economic systems} \and
      {89.75.Fb}{Structures and organization in complex systems} \and
      {89.75.Hc}{Networks and genealogical trees} \and
      {89.75.Da}{Systems obeying scaling laws} }
} 
\maketitle

\section{Introduction}

A social network is a set of people, each of whom is acquainted with
some subset of the others. In such a network the nodes (or vertices) 
represent people joined by edges denoting acquaintance or
co\-lla\-bo\-ra\-tion. Empirical data of social
networks include networks of scientific co\-lla\-bo\-ra\-tion\cite{NewmanPNAS},
of film actor co\-lla\-bo\-ra\-tions\cite{strogatznat}, friend\-ship
networks\cite{Amaral} among some others\cite{NewmanSIAM}.
One kind of social network is the 
network of sexual contacts\cite{nature93,Liljeros,UK,Latora}, 
where connections link those persons (agents) that have had 
sexual contact with each other. The empirical investigation 
of such  networks are of great interest because, 
e.g.~the topological features of sexual
partners distributions help to explain why persons can have the same
number of sexual partners and yet being at distinct risk levels of
contracting HIV~\cite{Klovdahl}. 

The simplest way to characterize the influence of each individual
on the network is through its degree $k$, the number
of other persons to whom the individual is connected. 
Sexual contact networks are usually addressed as an example of
scale-free networks\cite{Liljeros,UK,Latora,barabasirev}, because its property
of having a tail in its degree distribution, which
is well fitted by a power-law $P(k) \sim k^{-\gamma}$, with
an exponent $\gamma$ between $2$ and $3$. 
However, another characteristic feature,
not taken into account, is that the small $k$-region, comprehending 
the small-k values varies slowly with $k$, deviating from the power-law. 
Moreover, the size of the small-$k$ region also increases in time, 
yielding rather different distributions when considering the 
number of partners during a one year period or during 
the entire life, e.g. for entire-life sexual contacts, the
degree distribution shows that at least half of the nodes have degree 
in the small-$k$ region~\cite{nature93,Liljeros}.
A model predicting all these different distributions shapes for 
different time spans is of crucial interest, because 
the transmission of diseases occur during the growth mechanism 
of the network. 

One of the main difficulties for validating a model of sexual
interactions is that typical network studies of sexual contacts  
involve the circulation of surveys, or anonymous questionnaires, 
and only the number of sexual partners of each interviewed person 
is known, not being possible to obtain information about the entire 
network, in order to calculate degree correlations, closed paths 
(cycles), or average distance between nodes. 
\begin{figure}[t]
\begin{center}
\includegraphics*[width=8.5cm]{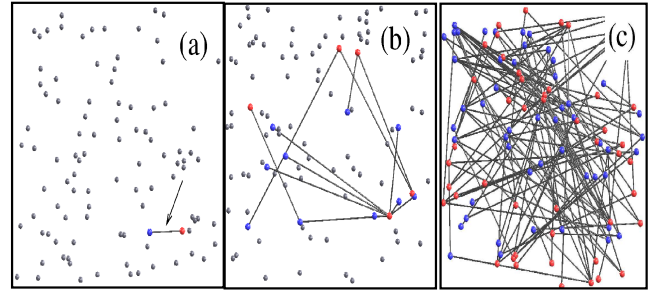}%
\end{center}
\caption{\protect
         Snapshots of the growing network of collisions in a low-density 
         gas with $N=100$, for
         {\bf (a)} $n=0.02N$,
         {\bf (b)} $n=0.15N$,
         {\bf (c)} $n=N$.
         In the on-line version two colors represent the two sexual genders
         and larger symbols emphasize those which belong to the network
         (linked agents).}
\label{fig1}
\end{figure}

In this work we propose a model of mobile agents in two dimensions
from which the network is build by keeping track of the collisions
between agents, representing the interactions among them.
In this way, the connections are a result not of some {\it a priori}
knowledge about the network structure but of some local dynamics of
the agents from which the complex networks emerge. 
Below, we show that this model is suitable to reproduce
sexual contact networks with degree distribution evolving in time,
and we validate the model using contact tracing studies from health 
laboratories, where the entire contact network is known. 
In this way, we are able to compare the number of cycles and average 
shortest path between nodes as well as compare the results with the
ones obtained with Barab\'{a}si-Albert scale-free networks\cite{AlbertScience}, 
which are well-known models, accepted for sexual 
networks\cite{Liljeros,UK,Latora}.
We start in Sec.~\ref{sec:model} by describing the model of mobile agents, 
and in Sec.~\ref{sec:sex} we apply it to reproduce the statistical features
of empirical networks of sexual contacts. Discussion and conclusions are 
given in Sec.~\ref{sec:conclusions}.
\section{Model of mobile agents for sexual interactions}
\label{sec:model}

The model introduced below is a sort of a granular 
system\cite{refpoeschel}, 
where $N$ particles with small diameter $d$ represent agents
randomly distributed in a two-dimensional system of linear size
$L\gg \sqrt{N}d$ (low density) 
and the basic ingredients are an increase of velocity when 
collisions produce sexual contacts, 
two genders for the agents (male and female), and $n/N$,
the fraction of agents that belong to the network, which constitutes
an implicit parameter for the resulting topology 
of the evolving the network. 

The system has periodic boundary conditions and is initialized as
follows: all agents have a randomly chosen gender, position and moving
direction with the same velocity modulus $\vert\vec{v}_0\vert$.
We mark one agent from which the network will be constructed. 
When the marked agent collides for the first time with another one of
the opposite gender, the corresponding collision is taken as the first 
connection of our network and its colliding partner is marked as the
second agent of the network (Fig.~\ref{fig1}a).   
Through time, more and more collisions occur, increasing the
size $n$ of the network (Fig.~\ref{fig1}b and \ref{fig1}c) till 
eventually all the agents composing the system are connected. 
\begin{figure}[htb]
\begin{center}
\includegraphics*[width=8.5cm,angle=0]{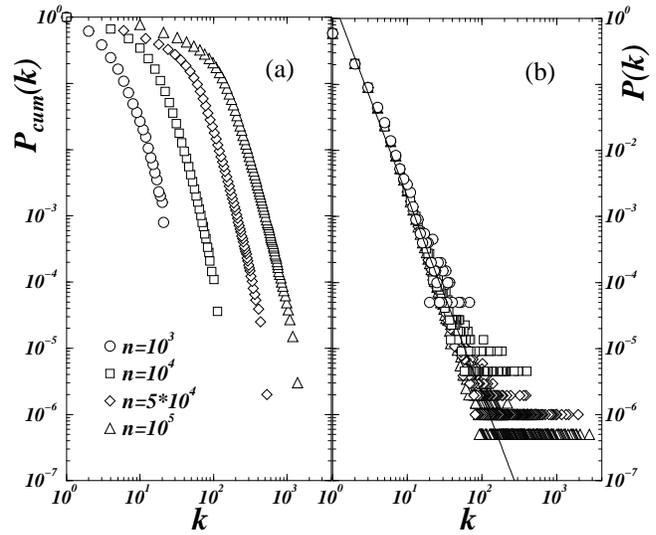}
\end{center}
\caption{\protect
         {\bf (a)} Cumulative distribution $P_{cum}(k)$ of the number 
         $k$ of partners among agents, when considering type-(i) and -(ii)
         interactions (see text) for
         $n=10^3$ (circles),
         $n=10^4$ (squares), $n=5\times 10^4$ 
         (diamonds) and $n=10^5$ (triangles).
         For the same parameter values {\bf (b)} shows
         a pure scale-free distribution, obtained when only type-(ii) 
         interactions form links.
         The solid line indicates the slope $\gamma=3$ of the scale-free 
         distribution.
         Here $\alpha=1$ and $N=320\times 320$.}
\label{fig2}
\end{figure}

Collisions between two agents take place whenever their distance is
equal to their diameter and the collision process is based on an
event-driven algorithm, i.e.~the simulation progresses by means of a
time ordered sequence of collision events and between collisions each
agent follows a ballistic tra\-jec\-to\-ry\cite{Rapaport}.
Since sexual interactions rely on the sociological
ob\-ser\-va\-tion\cite{Laumann} that individuals with a larger number of
partners are more likely to get new partners, we choose a collision
rule where the velocity of each agent increases with the number $k$ of
sexual partners. 
The larger the velocity one agent has the more likely it is to collide.  
Moreover, contrary to collision interactions where velocity direction
is completely deterministic\cite{ben-Avraham}, here the moving
directions after collisions are randomly selected, since in general,
sexual interactions do not determine the direction towards which each
agent will be moving afterwards. Therefore, momentum is {\it not}
conserved. 

Re\-gar\-ding these observations our collision rule for sexual
interactions reads
\begin{equation}
\vec{v}(k_i)=(k_i^{\alpha}+\vert\vec{v}_0(i)\vert)\vec{\omega} ,
\label{velo}
\end{equation}
where $k_i$ is the total number of sexual partners of agent $i$,
exponent $\alpha$ is a real positive parameter,
$\vec{\omega}=(\vec{e}_x\cos{\theta}+\vec{e}_y\sin{\theta})$ with 
$\theta$ a random angle and $\vec{e_x}$ and $\vec{e_y}$ are unit vectors.
Collisions which do not correspond to sexual interactions only change 
the direction of motion.

Collisions corresponding to sexual interactions, i.e.~with a velocity 
update as in Eq.~(\ref{velo}), are the only ones which produce links,
and occur in two possible situations: 
(i) between two agents which already belong to the network,
i.e.~between two sexually initiated agents and 
(ii) when one of such agents finds a non-connected (sexually non-initiated) 
agent.
For simplicity, we do not take into account sexual interactions between two
non-connected agents, and therefore our network is connected
(see the discussion in Sec.~\ref{sec:conclusions}).
\begin{figure}[htb]
\begin{center}
\includegraphics*[width=8.5cm,angle=0]{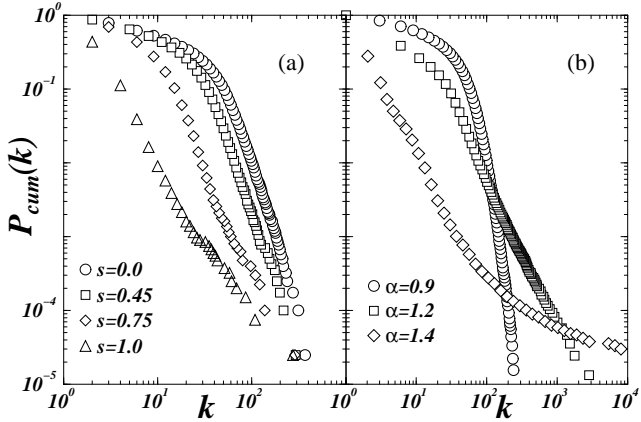}
\end{center}
\caption{\protect
         Cumulative distributions, when varying
         {\bf (a)} a parameter $s$ of selectivity which interpolates between 
         Figs.\ref{fig2}a and \ref{fig2}b (see text) for $\alpha=1$ and
         {\bf (b)} the exponent $\alpha$ in the update velocity rule,
         Eq.~(\ref{velo}), for $s=0$. Here the same stage of growth is 
         considered, namely $n=5\times 10^4=0.5N$.}
\label{fig3}
\end{figure}

When interactions of type (i) and (ii) occur, both the distribution tail 
and the small-$k$ region are observed, as shown by the cumulative distribution
$P_{cum}(k)$ in Fig.~\ref{fig2}a. 
Here, we use a system of $N=320\times 320$ agents with $\rho=0.02$, $\alpha=1$ and 
distributions are plotted for different stages of the network growth, namely
$n=10^3$,
$n=10^4$,
$n=5\times 10^4$ and 
$n=10^5\sim N$.
As one sees, the exponent of the power-law tail and the transition
between the tail and the small-$k$ region increase during the growth
process.
These features appear due to the fact that at later stages most of the
collisions occur between already connected agents. 
Consequently, the average number of partners increases as well.

If one considers only type-(ii) sexual contacts, the system reproduces
a stationary scale-free network, as shown in Fig.~\ref{fig2}b. 
In this case the average number of partners, defined as\cite{eames02}
$\langle k\rangle = k_{min}(\gamma-1)/(\gamma-2)$ with $k_{min}$ the minimum 
number of partners, is always $2$ ($k_{min}=1$ and $\gamma=3$).
As we show below, while empirical data of sexual contacts over large
periods have distributions like the ones for regime (i)+(ii), data for
shorter periods ($1-10$ years) are scale-free (only (ii)).
\begin{figure}[b]
\begin{center}
\includegraphics*[width=8.5cm,angle=0]{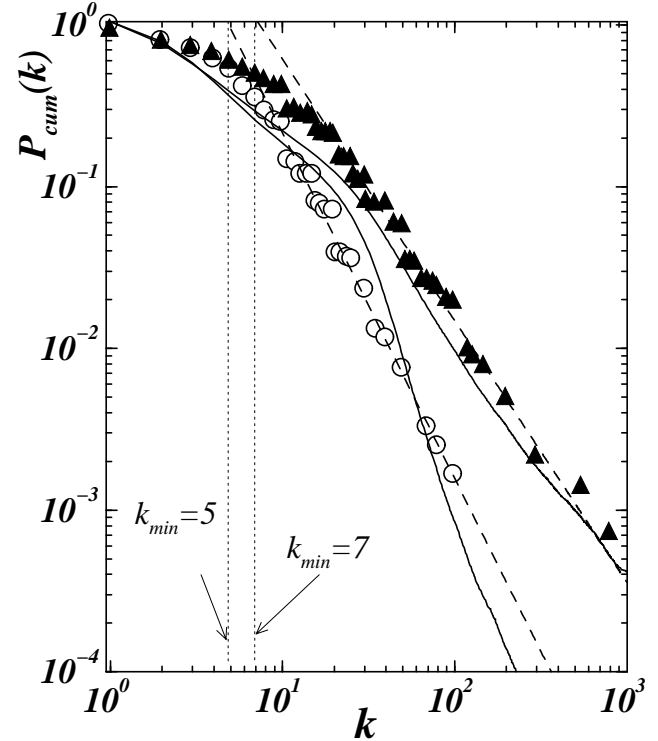}
\end{center}
\caption{\protect
         Cumulative distribution of sexual partners in a network 
         of heterosexual contacts extracted from Ref.~\cite{Liljeros}, 
         where male (triangles) and females (circles) distributions are
         plotted separately, with a total of $2810$ persons.
         Solid lines indicate the simulations when plotting the distributions 
         at the same stage $n=0.2N$, starting with a population composed by 
         $58\%$ of females and $42\%$ of males. 
         Here $N=10^5$, $s=0$}
\label{fig4}
\end{figure}
\begin{figure}[htb]
\begin{center}
\includegraphics*[width=8.5cm,angle=0]{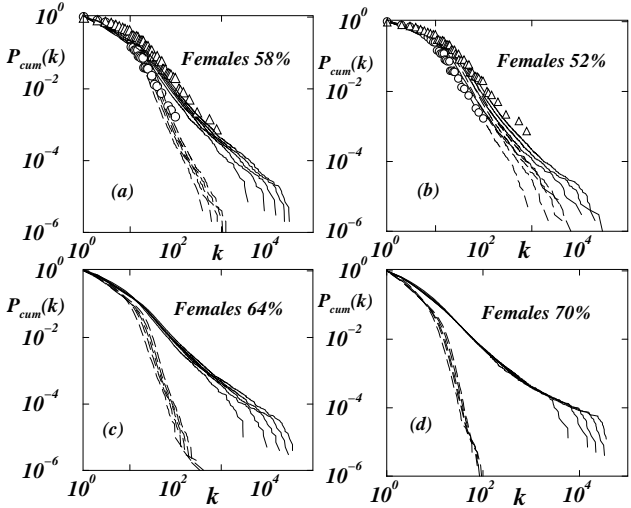}
\end{center}
\caption{\protect         
         Distributions of sexual partners 
         in a network of heterosexual contacts using different
         amounts of females:
         {\bf (a)} $58\%$ as in Fig.~\ref{fig4},
         {\bf (b)} $52\%$,
         {\bf (c)} $64\%$ and
         {\bf (d)} $70\%$.
         In each plot solid and dashed lines indicate the cumulative  
         distributions of males and females respectively, for
         five different realizations.
         Clearly, the exponent of the power-law tail of the
         distributions decreases when the percentage of
         females or males increases (see dotted lines).
         Same conditions as in Fig.~\ref{fig4} were used.}
\label{fig5}
\end{figure}
\begin{figure}[htb]
\begin{center}
\includegraphics*[width=8.5cm,angle=0]{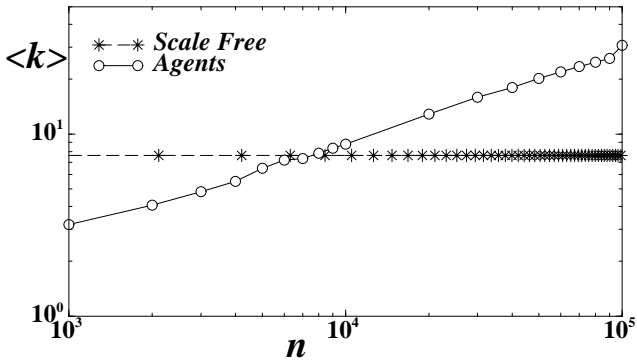}
\end{center}
\caption{\protect
        Comparing the average number of partners for scale-free networks
        (stars) and the agent model (circles). For the scale-free
        network $k_{min}=4$ and for agent model $N=520\times 520$,
        $s=0$.}
\label{fig6}
\end{figure}

With our model one can easily interpolate between both interaction
regimes, (i)+(ii) and (ii), by introducing a parameter $s$ of
`selectivity', defined as the probability that sexual initiated agents 
in case of collision with another initiated agent, have no sexual
contact.
Physically, this selectivity accounts for the intrinsic ability that a
node has to select from all its contacts (collisions) the ones which
are sexual. These intrinsic abilities were already used in other contexts,
e.g.~as a new mechanism leading to scale-free networks in cases where the 
power-law degree distribution is neither related to dynamical properties 
nor to preferential attachment\cite{Caldarelli}. 
For $s=0$ one obtains the two regions illustrated in Fig.~\ref{fig2}a,
namely the small-$k$ region and the power-law tail,
while for $s=1$ one obtains the pure scale-free topology illustrated in
Fig.~\ref{fig2}b.
In Fig.~\ref{fig3}a, we show the crossover between these two regimes. 

The shape of the cumulative distributions is also sensible to the
exponent $\alpha$ in the update velocity rule, Eq.~(\ref{velo}),  
as shown in Fig.~\ref{fig3}b. 
While for small values of $\alpha\lesssim 1$ one gets an
exponential-like distribution, for $\alpha\gtrsim 1.4$ the
distribution shows that a few nodes make most of the
connections. Henceforth, we fix $\alpha=1.2$.

Having described the model of mobile agents we proceed in the next section
of a specific application, i.e. modeling empirical networks of sexual
contacts. 
\section{Reproducing networks of sexual contacts}
\label{sec:sex}

In this Section we will show that,
by properly choosing the parameter values in our model, 
one can reproduce real data distributions of sexual contact networks.
In Fig.~\ref{fig4} the cumulative distributions of a real
contact network\cite{Liljeros} are shown for females (circles) and
males (triangles) separately, based on empirical data from $2810$
persons in a Swedish survey of sexual behavior.
The solid lines in Fig.~\ref{fig4} indicate the simulated distributions.
The simulated power-law tails have exponents
$\gamma_m=2.4$ and $\gamma_f=4.0$ 
for males and females respectively,
compared with the empirical data $\gamma_m=2.6 \pm 0.3$ and
$\gamma_f=3.1 \pm 0.3$\cite{Liljeros}.
To stress that, while the power-law tails are also well fitted by 
distributions obtained with scale-free networks (dashed lines in 
Fig.~\ref{fig4}), these  
distributions have a minimum number of connections (partners) of $k_{min}=5$ 
for females and $k_{min}=7$ for males, contrary to the real value $k_{min}=1$ 
also reproduced with our agent model. 
In fact, the model of mobile agents takes into account not only the power-law 
tail of these distributions, but also the small-$k$ region which comprehends 
the significant amount of individuals having only a few sexual partners
($k\gtrsim 1$).

Is important to note that in order to reproduce the difference 
in the exponents is necessary to have $58\%$ females and $42\%$ males, 
which is far from the expected difference among number of females
and males in typical human populations, with ratios of
females:males of the order of $1.1$.
The difference in the exponents of the distributions tails for
males and females separately, present in the data of sexual surveys, 
has generated much controversy and is often considered due to a bias 
either of sampling or honest reporting (see Ref.~\cite{nature93} and 
references therein). The exponents $\gamma_m$ and $\gamma_f$ on a bipartite
network are expected to be nearly the same when the percentage 
of males and females  are similar, as shown in 
Fig.~\ref{fig5}. In each plot five different realizations are shown
for males (solid lines) and females (dashed lines).
In Fig.~\ref{fig5}a, we plot the results for the same conditions as in 
Fig.~\ref{fig4} ($58\%$ females and $42\%$ males).
Taking the average of the five curves for each gender yields the
curves shown in Fig.~\ref{fig4}.
Figures \ref{fig5}b-d show the distributions obtained for other 
percentages illustrating that when decreasing the difference 
in the ratio females:males difference in the exponents dissappear.
\begin{figure}[htb]
\begin{center}
\includegraphics*[width=4.28cm,angle=0]{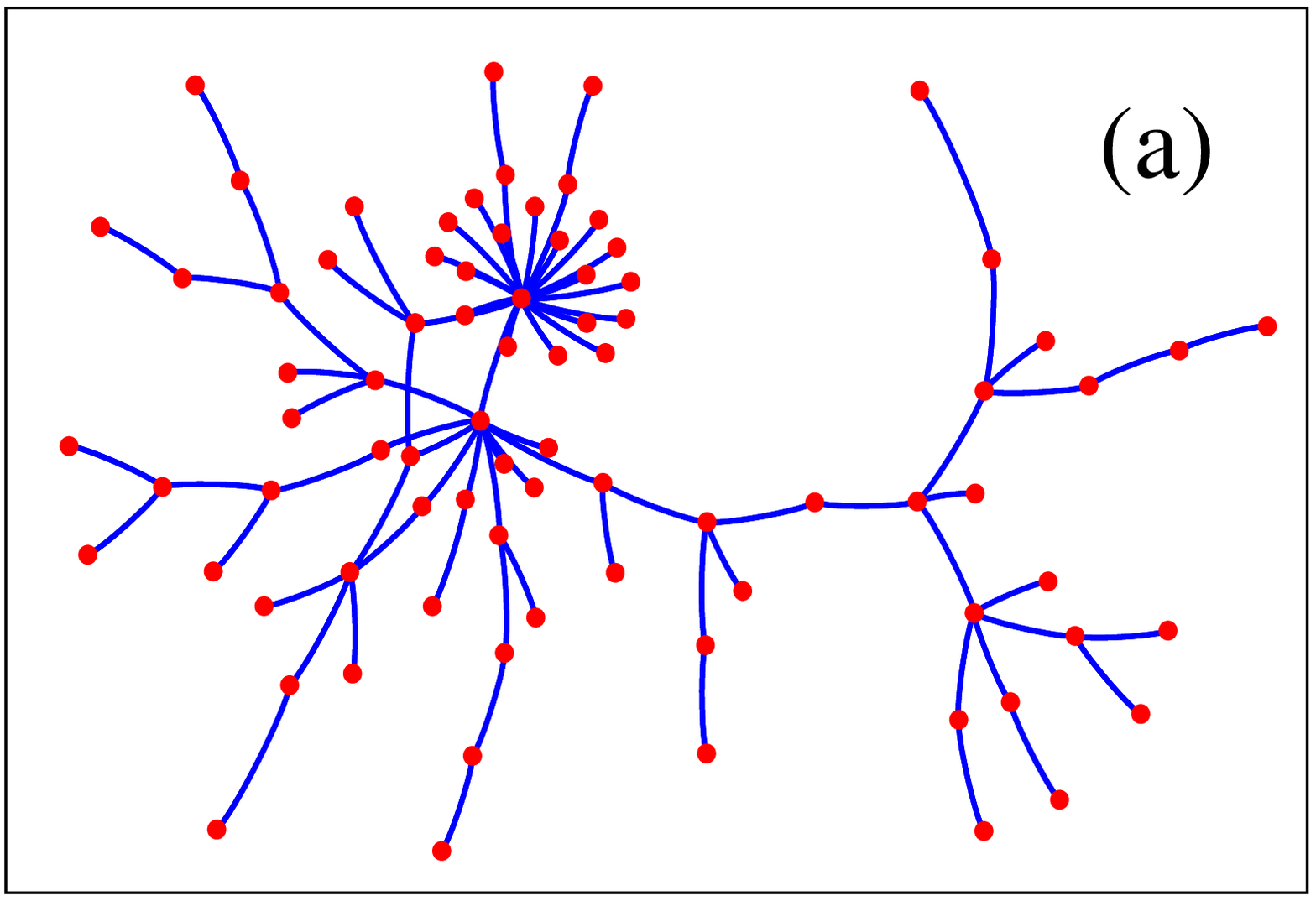}%
\includegraphics*[width=4.2cm,angle=0]{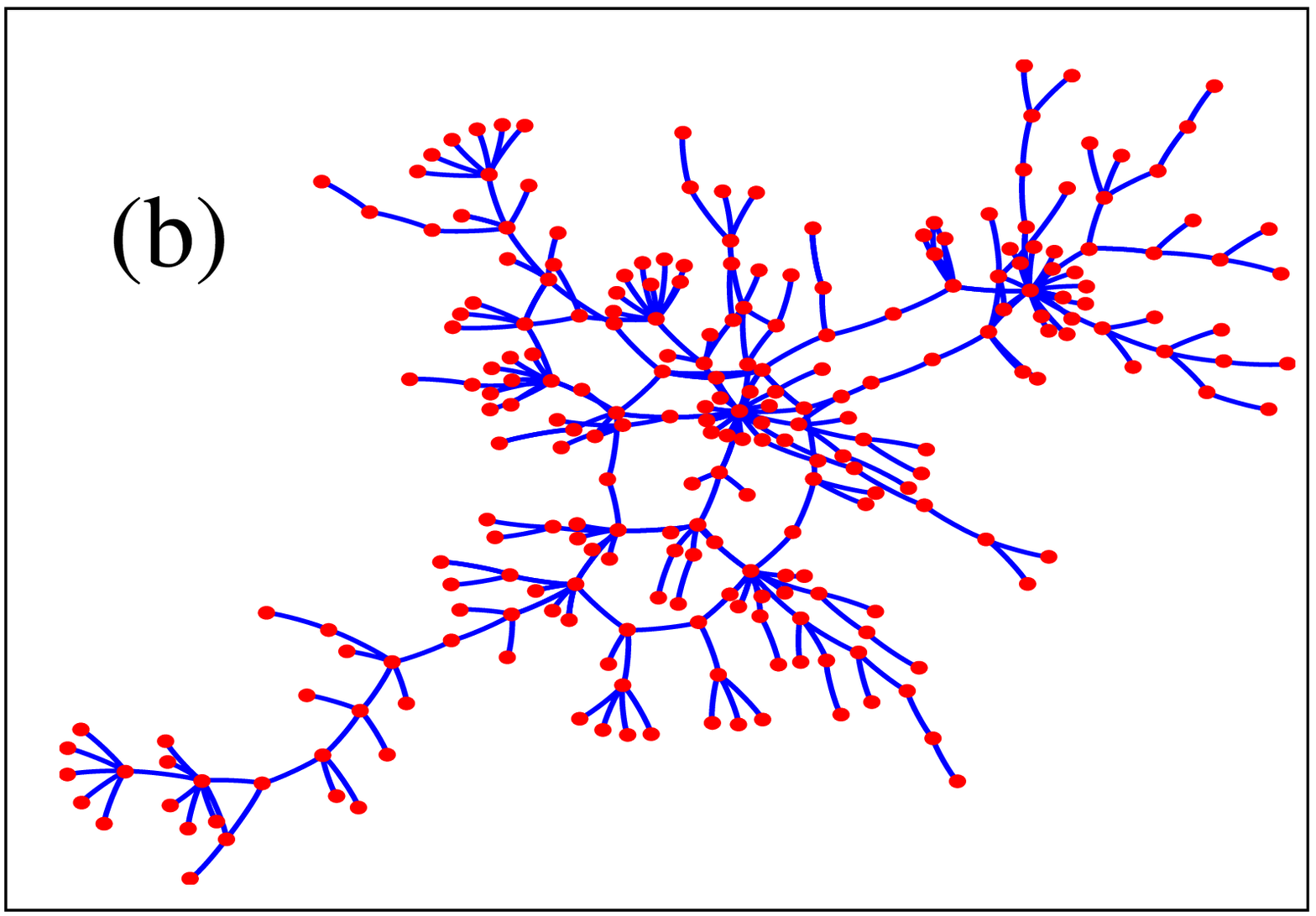}
\end{center}
\caption{\protect
         Sketch of two real sexual contact networks having
         {\bf (a)} only heterosexual contacts ($N=82$ nodes and
         $L=84$ connections) and
         {\bf (b)} homosexual contacts ($N=250$ nodes and $L=266$
         connections). 
         While in the homosexual network triangles and squares appear,
         in the heterosexual network triangles are absent (see
         Table \ref{tab1}).} 
\label{fig7}
\end{figure}
\begin{figure}[t]
\begin{center}
\includegraphics*[width=8.5cm,angle=0]{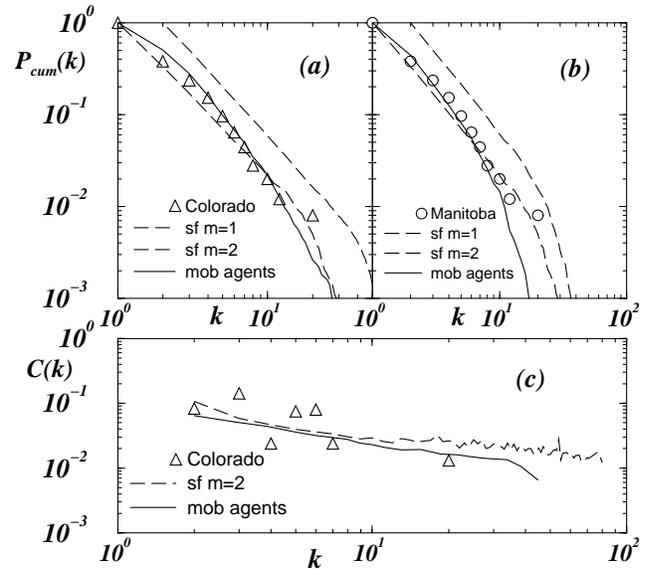}
\end{center}
\caption{\protect
         {\bf(a)} Cumulative degree distribution of a homosexual contact
         network\cite{cospring2} with $n=250$ (triangles). {\bf(b)} Cumulative 
         degree distribution of a heterosexual contact
         network\cite{cospring1} with $n=82$ (triangles). Each case
         is compared with the average degree distribution over $20$
         iterations, for the $BA$ scale-free model
         (dashed line) with $k_{min}=1$, and $k_min=2$ and 
         with our agent model (solid line) with $s=0.7$. 
         {\bf (c)} Cluster coefficient for the homosexual network
         empirical data (triangles), the agent model (solid
         line) and the $sf$ model $k_{min}=2$. The scale-free 
         $k_{min}=1$ yields $C(k)=0$ (not shown).}
\label{fig8}
\end{figure}
\begin{table}[htb]
\centering
\begin{tabular}{|c||c|c|c|c|c|}
\hline
  & $N$ & $L$ & $T$ & $Q$ & $C$ \\ 
\hline\hline
Heterosexual & $82$ & $84$ & $0$  & $2$ & $0$ \\ 
  & & & & & \\ \hline 
Homosexual \phantom{.}  & $250$ & $266$ & $11$ & $6$ & \ $0.02980$ \\ 
  & & & & & \\\hline\hline 
Heterosexual  & $82$ & $83.63$ & $0$ & $1.45$ & $0$ \\ 
\footnotesize{(Agent Model)}  & & & & & \\\hline 
Homosexual\  & $250$ & $287.03$ & $8.23$ & $10.52$ & $0.02302$ \\ 
\footnotesize{(Agent Model)}  & & & & & \\\hline\hline 
Heterosexual  & $82$ & $162$ & $0$ & $159.72$ & $0$ \\ 
\footnotesize{(Scale-free)}  & & & & & \\\hline 
Homosexual\  & $250$ & $498$ & $45.28$ & $256.79$ & $0.08170$ \\ 
\footnotesize{(Scale-free)}  & & & & & \\\hline 
\end{tabular}
\caption{\protect
         Clustering coefficients and cycles in two real networks
         of sexual contacts (top), one
         where all contacts are heterosexual and
         another with homosexual contacts.
         In each case one indicates the values of the number $N$ of
         nodes, the number $L$ of connections, the number $T$ of
         triangles, the number $Q$ of squares and the average clustering
         coefficient $C$.
         The values of these quantities are also indicated for networks
         constructed with the agent model and for scale-free networks with
         $k_{min}=2$, note that for $k_{min}=1$, L=81 and L=249, respectively
         and there are not cycles.}
\label{tab1}
\end{table}
A characteristic feature of our model is that the average
number $\langle k\rangle$ of partners increases as the network
grows, which is natural characteristic expected to occur in real sexual
networks according to the observed differences in the shape
of the degree distribution for yearly and
entire-life reports of number of sexual partners
\cite{Liljeros,UK}. This feature is not observed in 
scale-free networks, as illustrated in Fig.~\ref{fig6}.
Of course, that this growth also indicates non-stationary
regimes, where $\langle k\rangle$ diverges with the network 
growth. In the next section we explain how to overcome this shortcoming.

We compare the model, with two empirical networks of sexual contacts.
One network is obtained from an empirical data set, composed solely
by heterosexual contacts among $n=82$ nodes, extracted at the Cadham
Provincial Laboratory (Manitoba, Canada) 
and is a 6-month block data~\cite{cospring1}
between November 1997 and May 1998 (Figure \ref{fig7}a sketches
this network). 
The other data set is the largest cluster with $n=250$ nodes in the
records of a contact tracing study~\cite{cospring2}, from 1985 to
1999, for HIV tests in Colorado Springs (USA), where most of the
registered contacts were homosexual (see Figure \ref{fig7}b).  

Figures \ref{fig8}(a)-(b) show the cumulative distribution of the number of
sexual partners for each of the empirical networks.
For both cases the agent model and scale-free networks with $k_{min}=1$
can reproduce the distribution of the number of partners. However, the agent
model with $s=0.7$ reproduces, as well, the clustering coefficient
distribution that we measure from the empirical network.

The clustering coefficient $C(i)$ of one agent is defined\cite{strogatznat} 
as the total number of triangular loops of connections passing through
it divided by the total number of connections $k_i$. Averaging $C(i)$ over
all nodes with $k_i$ neighbors yields the clustering coefficient distribution
$C(k)$.   
While for the scale-free graph which better reproduces these empirical data, 
the clustering coefficient is zero, our agent 
model yields a distribution which resembles the one observed 
in the real network (Fig.~\ref{fig8}c). 
This feature is due to the co-existence of a tree-like
substructure and closed paths (see Figs.~\ref{fig7}b).

For both heterosexual and homosexual networks of sexual contacts, 
the model of mobile agents reproduces other important statistical
features, namely the average clustering coefficient $C$ and the number of
loops of a given order. Table \ref{tab1} indicates
the number $T$ of triangles (loops composed by three edges), 
the number $Q$ of squares (loops with four edges) and the 
average clustering coefficients $C$ given by\cite{strogatznat} the average 
of $C(i)$ over the entire network.

When using the agent model with the same number $N$ of nodes as in the real 
networks we obtain similar results for $L$, $T$, $Q$ and $C$, as shown in Table
\ref{tab1} (middle), where values represent averages over samples of $100$
realizations. For the heterosexual network there are no triangles due to
the bipartite nature of the network.
At the bottom of Table \ref{tab1} we also show the values obtained with 
scale-free networks, for both cases of one and two genders, whose minimum
number of connections was chosen to be $k_{min}=2$, for which the clustering
coefficient distributions are as close as possible from 
the distributions of the real networks.
Clearly, the agent model not only yields clustering coefficient values
much closer to the ones measured in the empirical data, but also does not
show the formation of huge amounts of loops (triangles and squares), a 
feature of scale-free networks which is not observed in empirical data.

\section{Discussion and conclusions}
\label{sec:conclusions}

In this paper we presented a new model for networks
of complex interactions, based on a granular system of mobile a\-gents
whose collision dynamics is governed by an efficient event-driven
algorithm and generate the links (contacts) between a\-gents. 
As a specific application, we showed that the dynamical rules for
interactions in sexual networks can be written as a velocity update
rule which is a function of a power $\alpha$ of the previous contacts
of each colliding agent.  
For suitable values of $\alpha$ and selectivity $s$, the model
not only reproduces empirical data of networks of sexual contacts but
also generates networks with similar topological features as the
real ones, a fact that is not observed when using standard scale-free 
networks of static nodes.

Furthermore, our model predicts that the growth mechanism of sexual
networks is not purely scale-free, due to interactions among internal
agents, having a mean number of partners which increases in time.
This should influence the predictions from models  of spreading 
of infections\cite{eames02}.
Our agent model offers a realistic approach to study the emergence of
complex networks of interactions in real systems, using only local
information for each agent, and may be well suited to study networks
in sociophysics, biophysics and chemical reactions, where interactions
depend on specific local dynamical behavior of the elementary agents
composing the network. 

While given promising results the model may be improved in two particular 
aspects. 
First, it should enable the convergence towards a stationary regime with 
a growth process starting with all possible collisions instead of one 
particular agents from which the network is constructed. 
Second, the dependence of the above results on the velocity rule in 
Eq.~(\ref{velo}) should be studied in detail, namely for the case of
constant velocity ($\alpha=0$).
Preliminary results have shown that the stationary regime is easily
obtained with the model above by introducing a simple aging scheme,
while by varying the parameter $\alpha$ one is able to reproduce
other non-trivial degree distributions.
Moreover, we introduced the selectivity parameter $s$ to select from 
all possibles social interactions (collisions) the ones which are of 
sexual nature.
Without introducing this selectivity, the model of mobile agents is able 
to reproduce other social networks of acquaintances.
These and other questions will be addressed elsewhere\cite{new}.

\section*{Acknowledgments}

The authors would like to thank Jason A.C. Gallas, Dietrich Stauffer,
Maya Pa\-czuski, Ram\'on Garc\'{\i}a-Rojo and Hans-J\"{o}rg Seybold 
for useful discussions. 
MCG thanks Deutscher Aka\-demischer Austausch Dienst (DAAD), Germany, and
PGL th\-anks Funda\c{c}\~ao para a Ci\^encia e a Tecnologia (FCT), Portugal, 
for financial support. 



\end{document}